\documentclass[a4paper,11pt]{article}
\pdfoutput=1 

\usepackage{jheppub} 

\usepackage[T1]{fontenc} 

\usepackage{float} 

\usepackage{amsmath}

\title{\boldmath Electroluminescence yield in pure krypton}

\author[a]{R.D.P. Mano,}
\author[a]{C.A.O. Henriques,}
\author[a]{F.D. Amaro}
\author[a,1]{and C.M.B. Monteiro\note{Corresponding author.}}

\affiliation[a]{LIBPhys-UC, Physics Department, University of Coimbra,\\Rua Larga, 3004-516 Coimbra, Portgal}

\emailAdd{cristinam@uc.pt}

\abstract{The krypton electroluminescence yield was studied, at room temperature, as a function of electric field in the gas scintillation gap. A large area avalanche photodiode has been used to allow the simultaneous detection of the electroluminescence pulses as well as the direct interaction of x-rays, the latter being used as a reference for the calculation of the number of charge carriers produced by the electroluminescence pulses and, thus, the determination of the number of photons impinging the photodiode. An amplification parameter of 113 photons per kV per drifting electron and a scintillation threshold of 2.7 Td ( 0.7 kV cm$^{-1}$ bar$^{-1}$ at 293 K ) was obtained, in good agreement with the simulation data reported in the literature. On the other hand, the ionisation threshold in krypton was found to be around 13.5 Td (3.4 kV cm$^{-1}$ bar$^{-1}$), less than what had been obtained by the most recent simulation work-package. The krypton amplification parameter is about 80\% and 140\% of those measured for xenon and argon, respectively. The electroluminescence yield in krypton is of great importance for modeling krypton-based double-phase or high-pressure gas detectors, which may be used in future rare event detection experiments.}

\begin{document} 
\maketitle
\flushbottom

\section{Introduction}
\label{sec:intro}

The electroluminescence yield of gaseous xenon and argon has been studied in detail, both experimentally (e.g. see \cite{1,2,3,4,5,6,7} and references therein) and through simulation tools \cite{8,9,10,11,12}. At present, the main drive for those studies is the ongoing development of dual-phase \cite{13,14,15,16,17,18,19} and high-pressure gaseous \cite{20,21,22,23} optical Time Projection Chambers (TPC), which make use of the secondary scintillation, - electroluminescence (EL) - processes in the gas for the amplification of the primary ionisation signals produced by radiation interaction inside the TPC active volume. The R\&D of such TPCs aims at application to Dark Matter search \cite{13,14,15,16,17} and to neutrino physics, such as neutrino oscillation \cite{18,19}, double beta decay \cite{20,21,22} and double electron capture \cite{24} detection. The physics behind these rare event detection experiments is of paramount importance in contemporary particle physics, nuclear physics and cosmology, justifying the enormous R\&D efforts carried out by the scientific community.

The radioactivity of $^{85}$Kr has been a drawback for the use of krypton in rare event detection experiments, being this gas the less studied one among the noble gases. To the best of our knowledge, the electroluminescence in krypton has only been studied by simulation \cite{8,10} and there haven’t been published any experimental results, up to now, to benchmark the simulation tools.  Nevertheless, there are two experiments that make use of krypton, namely the measurement of the double electron capture in $^{78}$Kr \cite{25,26,27,28} and the search for solar hadronic axions emitted in the $M$1 transition of $^{83}$Kr nuclei \cite{29,30,31,32}. Moreover, $^{83}$Kr has also been proposed for inelastic dark matter search \cite{33}. The enrichment of a given isotope of a noble gas is, nowadays, a matured technique, not significantly expensive, allowing for the reduction of the radioactive isotope to tolerable levels for a particular experiment.

The double-electron capture half-life is an important benchmark for nuclear structure models \cite{34,35,36,37,38}, providing vital experimental constraints. In addition, it presents a significant step in the search for neutrinoless double electron capture. The latter can complement the search for neutrinoless double beta decay. Both would unveil the Majorana nature of the neutrino, access the absolute neutrino mass and contribute to understand the dominance of matter over antimatter by means of leptogenesis. On the other hand, axions and axion-like particles are potential candidates for the constituent particles of dark matter, being the main reason for extensive axion searches, e.g. see \cite{39} and references therein for detailed theoretical and experimental reviews.

The referred to above rare event search experiments, having krypton as the target, have been carried out with gas proportional counters using enriched krypton \cite{25,26,27,28,29,30,31,32}. Nevertheless, optical TPCs deliver higher gains with reduced electronic noise and overly improved energy resolution when compared to proportional counters \cite{40,41,42,43}. In addition, the use of a 2D-readout for the EL signal allows the reconstruction of the topology of the ionisation event \cite{42,43} in a more effective way than the complex analysis of the wave form associated to the ionisation events in the above-mentioned proportional counters. Therefore, the use of optical TPCs will allow larger sensitive volumes, better event discrimination and a more effective background reduction than the present proportional counters, hence having the potential for improved sensitivity and accuracy.

Having this in mind, in this work we present an experimental study of the electroluminescence yield of pure Kr and compare the obtained results with those attained from simulation studies \cite{8,9,10}. The setup is described in section \ref{sec:2 setup}, and in section \ref{sec:3 method} we discuss the methodology that was followed in order to obtain the absolute EL yield, while in section \ref{sec:4 results} we present the obtained results and corresponding discussion, summarising the main conclusions in section \ref{sec:5 concl}.

\section{Experimental setup}
\label{sec:2 setup}

\begin{figure}[H]
\centering
\includegraphics{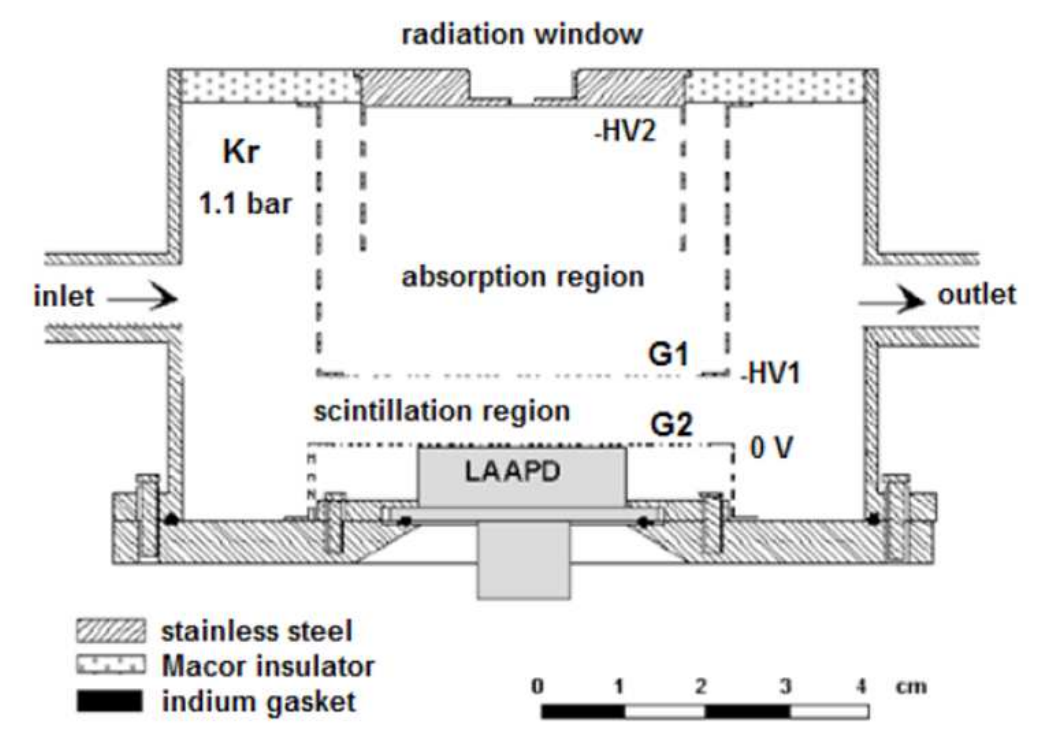}
\caption{\label{Fig1} Schematic of the GPSC with a large-area APD as the photosensor.}
\end{figure}

In this work, we used a gas proportional scintillation counter (GPSC) \cite{40}, Fig.\ref{Fig1}, 

\noindent irradiated with a 1.5-mm collimated beam of 5.9-keV x-rays emitted from a $^{55}$Fe radioactive source, having a Cr filter to absorb the most part of the 6.4-keV Mn $K_{\beta}$ line.
The primary electron cloud resulting from the x-ray interactions in the absorption region are driven towards the scintillation region by a weak electric field, with intensity below the gas excitation threshold. The intensity of the electric field in the scintillation region is set above the gas excitation threshold, but below the ionisation threshold, to prevent electron multiplication. Upon crossing the scintillation region, the electrons are accelerated by the electric field, gaining enough energy to excite the gas media by electron impact, leading to an electroluminescence pulse with a large number of VUV photons as a result of the gas de-excitation processes. The number of VUV photons is proportional to the number of primary electrons crossing the scintillation region and, thus, to the incident x-ray energy. The detector response to x-rays has been studied in detail for Xe \cite{46}, Ar \cite{47} and Kr \cite{48} gas fillings.

The GPSC depicted in Fig.1 has a 2.5-cm deep absorption region, a 0.9-cm deep scintillation region and is filled with Kr at a pressure of 1.1 bar, continuously purified through St707 SAES getters \cite{44}. G1 and G2 are meshes, made out of stainless steel wires, 80-$\mu$m in diameter with 900-$\mu$m spacing. The radiation window holder and its focusing electrode are made of stainless steel and are kept at negative voltage, while the stainless steel G2-holder and the detector body are maintained at ground potential. The voltage difference between the radiation window and G1 determines the electric field in the absorption region, while the voltage of G1 determines the electric field in the scintillation region. A Macor piece isolates the radiation window holder, the G1 holder and its feedthrough, being vacuum-sealed onto the stainless steel using a low-vapour pressure epoxy.
The electroluminescence pulses are readout by a VUV-sensitive silicon large-area avalanche photodiode (LAAPD) \cite{45}, having a 16-mm diameter active area. The LAAPD is vacuum-sealed by compressing its enclosure against the detector bottom plate, using an indium ring. The LAAPD signals are fed through a low-noise, 1.5 V/pC, charge pre-amplifier followed by an amplifier with 2 $\mu$s shaping time and are pulse-height analysed with a multi-channel analyser (MCA).

\section{Absolute electroluminescence yield measurement methodology}
\label{sec:3 method}

\begin{figure}[H]
    \centering
    \includegraphics[scale=0.6]{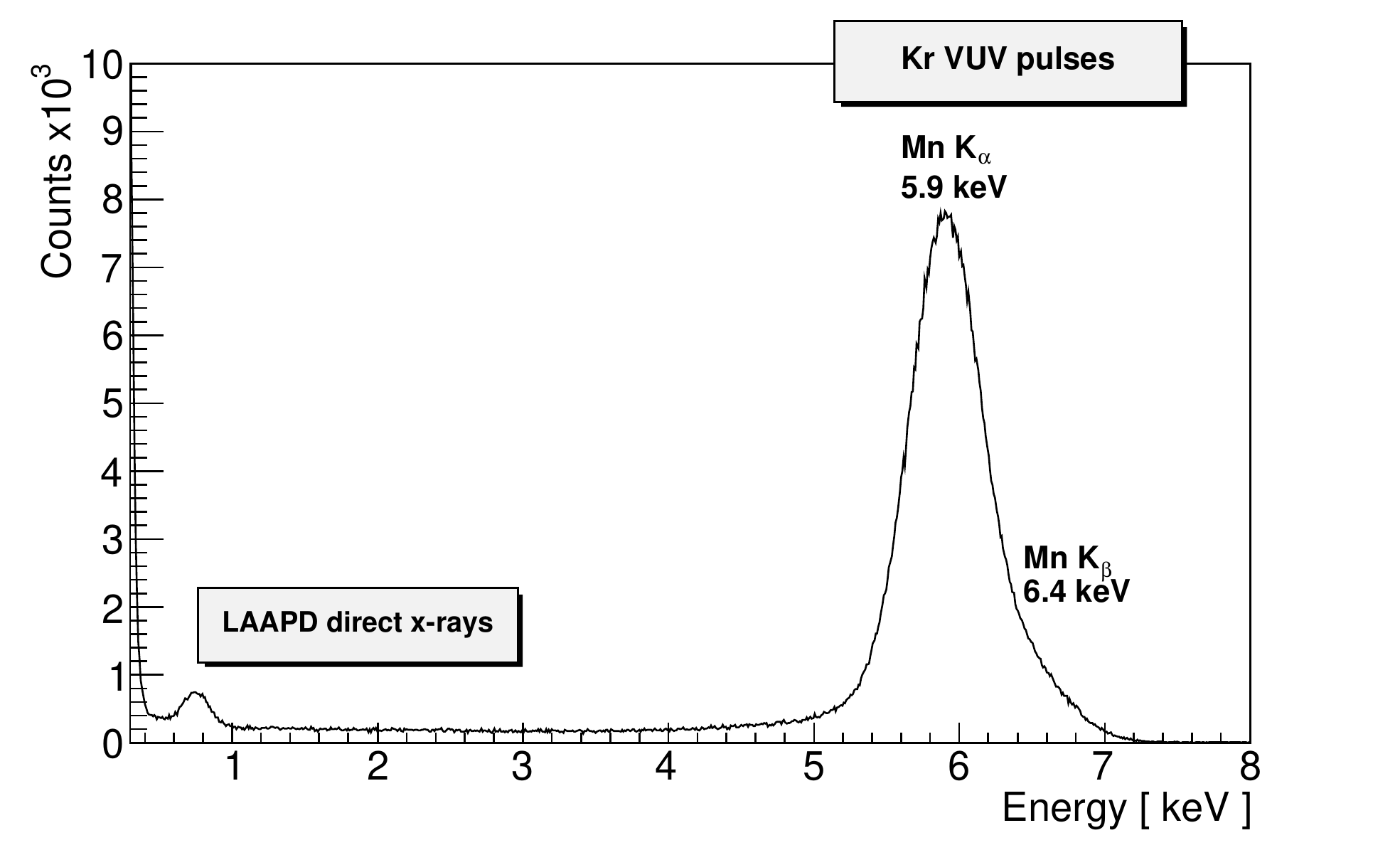}
    \caption{\label{Fig2} Typical pulse-height distribution obtained for 5.9-keV x-rays and electric field intensities of 0.34 and 3.4 kV cm$^{-1}$ bar$^{-1}$ in the absorption and scintillation region, respectively. The LAAPD bias voltage was 1840 V, corresponding to a gain of $\sim$ 150.}
\end{figure}

Most of the 5.9-kev x-rays interact in the absorption region producing, in the LAAPD, signals of large amplitude as a result of the electroluminescence. Nevertheless, a small fraction of the 5.9 keV x-rays are transmitted through the gas and interact directly in the LAAPD producing signals with lower amplitude when compared to those resulting from the x-ray interactions in the gas. Fig.\ref{Fig2} depicts a typical pulse-height distribution of the signals at the LAAPD output, obtained when irradiating the detector with 5.9-keV x-rays. It includes the Kr electroluminescence peak, more intense and in the high-amplitude region, the peak resulting from the direct interactions of the x-rays in the LAAPD, much less intense and in the low-amplitude region, and the electronic noise tail in the low-amplitude limit. While the amplitude of the electroluminescence peak depends on both scintillation region- and LAAPD-biasing, the amplitude of the events resulting from direct x-ray interaction in the LAAPD just depends on the LAAPD-biasing. In addition, the latter peak is present even for a null electric field in the scintillation region and/or when the detector is under vacuum. For pulse amplitude measurements, the pulse-height distributions were fit to Gaussian functions superimposed on a linear background, from which the Gaussian centroids were determined.

The presence of the peak of direct x-ray interactions in the LAAPD is of utmost importance, as the average number of charge carriers produced in the silicon wafer by the x-ray interactions, $N_{RX}$ is well known and it is used as a reference for the electroluminescence peak. Comparing both pulse-heights, a ratio can be found between the pulse amplitudes resulting from the electroluminescence and from the direct x-ray absorption in the LAAPD. This ratio allows a direct quantification of the number of charge carriers produced in the LAAPD by the electroluminescence pulse and, thus, the number of VUV-photons impinging the LAAPD, given its quantum efficiency. The concurrent detection of the light pulses and the x-rays in the photosensor under the same conditions in the same setup allows a straight forward measurement of the number of photons impinging on it. This method has been used for measuring Xe and Ar electroluminescence yield in uniform electric fields \cite{1,2} and in electron avalanches of GEMs, THGEMs and Micromegas micropatterned structures \cite{49,50}.
The number of charge carriers produced in the LAAPD by the electroluminescence pulse is:

\begin{equation}
\label{eq:1}
N_{VUV} =\left(\frac{A_{EL}}{A_{RX}}\right)N_{RX} 
\end{equation}

\noindent being $A_{RX}$ and $A_{EL}$ the amplitude of the peaks resulting from the direct x-ray interactions in the LAAPD and from the scintillation produced by the x-ray interactions in the gas, read in the MCA. The non-linear response of the LAAPD to 5.9-keV x-rays, e.g. see Fig.20 of \cite{51}, was considered as the ratio $A_{EL}$/$A_{RX}$ corrected for this effect. A factor of 0.94 was used in this correction, for a LAAPD biasing voltage of 1840 V used throughout this work, which corresponds to a LAAPD gain about 150. Knowing the quantum efficiency, QE, of the LAAPD and the optical transparency of G2, \textit{T}, and the fraction of the average solid angle, $\Omega_f = \Omega/4\pi$ subtended by the active area of the photosensor relative to the primary electron path, the total number of photons produced in the electroluminescence pulse is obtained by:

\begin{equation}
\label{eq:2}
N_{total,VUV} =\left(\frac{A_{EL}}{A_{RX}}\right)\frac{N_{RX}}{QE\times T \times\Omega_f} 
\end{equation}

The electroluminescence yield is defined as the number of photons produced per drifting primary electron per unit path length:

\begin{equation}
\label{eq:3}
Y = \left(\frac{A_{EL}}{A_{RX}}\right)\frac{N_{RX}}{QE\times T \times\Omega_f \times N_e \times d}
\end{equation}

\noindent where $N_e$ is the number of primary electrons produced in Kr by a 5.9 kev x-ray interaction and $d$ is the scintillation region depth. 

As the w-value in silicon is 3.62 eV, e.g. \cite{52} and references therein, the average number of free electrons produced in the LAAPD by the full absorption of the 5.9-keV x-rays is $N_{RX}$ = 1.63$\times$10$^3$ electrons. The w-value in Kr is 24.2 eV \cite{53}, thus being $N_e$ = 244 electrons. The optical transmission of G2 mesh is T=83$\%$ and the fraction of the average solid angle has been computed by Monte Carlo simulation \cite{54} to be $\Omega_f$=0.215. 

At atmospheric pressures, the electroluminescence of Kr consists of a narrow line peaking at 148 nm with 5 nm FWHM \cite{55}, called second continuum, being the emissions in the visible and in the IR regions below few percent in comparison with that in the VUV range \cite{55,56}, thus considering its contribution negligible. The processes leading to emission in the second continuum can be schematized as

\

\noindent $e^- + Kr \rightarrow e^- + Kr^*$,

\vspace{0.1cm}

\noindent $Kr^* + 2Kr \rightarrow Kr^*_2 + Kr$,

\vspace{0.1cm}

\noindent $Kr^*_2 \rightarrow 2Kr + h\nu$.

\

The electron impact with Kr atoms induces excited atoms, which through three-body collisions creates excited excimers, $Kr^*_2$, that decay emitting one VUV photon, $h\nu$. It corresponds to transitions of the singlet and triplet bound molecular states, from vibrationally relaxed levels, to the repulsive ground state.

A LAAPD QE was measured to be 0.90 for Kr EL \cite{57}. According to the manufacturer, the LAAPD fabrication technology is well established, and quite good reproducibility is obtained and it is expected that the behaviour observed for individual LAAPDs is representative for any of these devices. In fact, \cite{58} have measured the relative QE of $\sim$ 600 LAAPDs and obtained an approximately Gaussian distribution with a FWHM of $\sim$ 0.1. Therefore, we have considered an uncertainty of $\pm$ 0.08 ($\sim$ 2$\sigma$) for the LAAPD QE, being this the major source of uncertainty in our measurements. 

\section{Experimental results and discussion}
\label{sec:4 results}

\begin{figure}[H]
    \centering
    \includegraphics[scale=0.45]{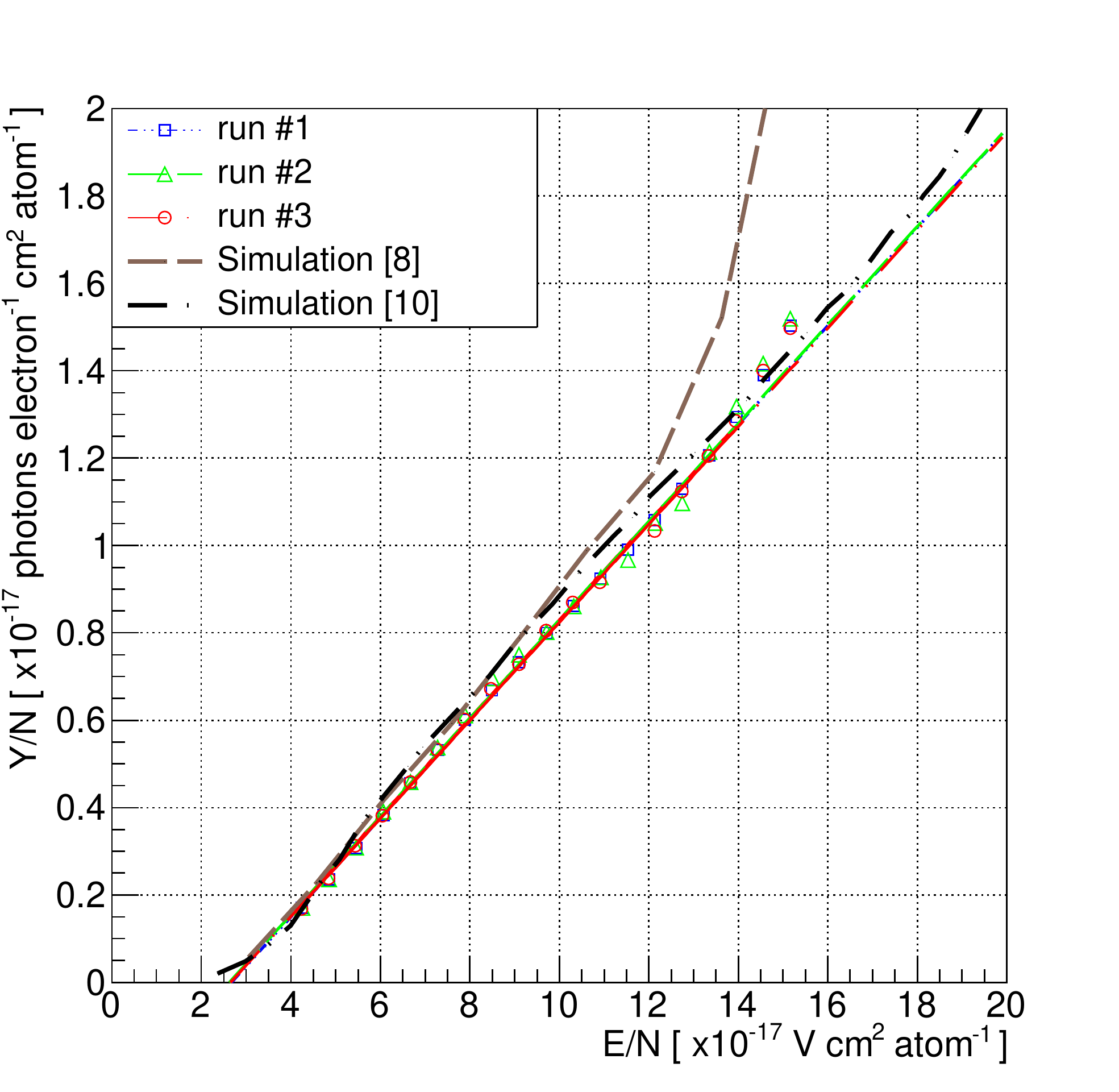}
    \caption{\label{Fig3} Krypton reduced electroluminescence yield as a function of reduced electric field for three different runs (this work) and the respective linear fit to the data below 14 Td, as well as for data obtained from Monte Carlo simulation \cite{8,10}.}
\end{figure}

In Fig.\ref{Fig3} we depict the reduced electroluminescence yield, Y/N, i.e. the electroluminescence yield divided by the number density of the gas, as a function of reduced electric field, E/N, in the scintillation region. The data was taken using a constant electric field of 0.36 kV/cm in the absorption region, while varying the electric field in the scintillation region. Three independent runs have been performed with several days of interval and with a room temperature between 25 and 26$^{\circ}$C, showing a good reproducibility of the experimental results. The reduced EL yield exhibits an approximately linear trend with the reduced electric field, a behavior similar to that of Ar and Xe and mixtures of Xe with He or with molecular additives \cite{1,2,3,4,5,6,7,8,9,10,11,12}. For each run, Fig.\ref{Fig3} also depicts a linear fit superimposed to the experimental data, excluding the two data points taken at the highest reduced electric field where secondary ionisation is already non-negligible. Simulation results from \cite{8,10} are also depicted for comparison. To the best of our knowledge, there are no other experimental or simulation results in the literature. 

The reduced electroluminescence yield can be approximately represented as 

\begin{equation}
\label{eq:4}
    \mathrm{Y/N \ (10^{-17} \ photons \ electron^{-1} \ cm^{2} \ atom^{-1}) = 0.113 \ E/N - 0.30}
\end{equation}

\noindent where E/N is given in Td (10$^{-17}$ V cm$^2$ atom$^{-1}$). This equation can also be represented as a function of pressure, at a given temperature used to convert the gas density into pressure. At room temperature Eq.\ref{eq:4} can be expressed as

\begin{equation}
\label{eq:5}
    \mathrm{Y/p \ (photons \ electron^{-1} \ cm^{-1} \ bar^{-1}) = 113 \ E/p - 74}
\end{equation}

The slope of the linear dependence denotes the scintillation amplification parameter, i.e., the number of photons produced per drifting electron and per volt, and is in good agreement with what has been obtained by Monte Carlo simulations for room temperature \cite{8,10}.  In addition, the excitation threshold for Kr, defined as the extrapolation of the linear trend to zero scintillation, 2.7 Td (0.7 kV cm$^{-1}$ bar$^{-1}$ at 293 K), is also in good agreement with the values presented in the literature \cite{8,10,59}. On the other hand, at a reduced electric field of 14 Td (3.5 kV cm$^{-1}$ bar$^{-1}$) the experimental data depart from the linear trend, denoting already the presence of a non-negligible amount of charge multiplication in the scintillation region. Therefore, from the experimental data we can conclude that the Kr ionisation threshold should be around 13.5 Td (3.4 kV cm$^{-1}$ bar$^{-1}$). This value is above that obtained from the MC simulation of \cite{8} and is lower than foreseen by the most recent simulation toolkit \cite{10}, demonstrating the importance of the data obtained in this work, for the development of future optical-TPCs based on Kr filling.

\begin{figure}[H]
    \centering
    \includegraphics[scale=0.45]{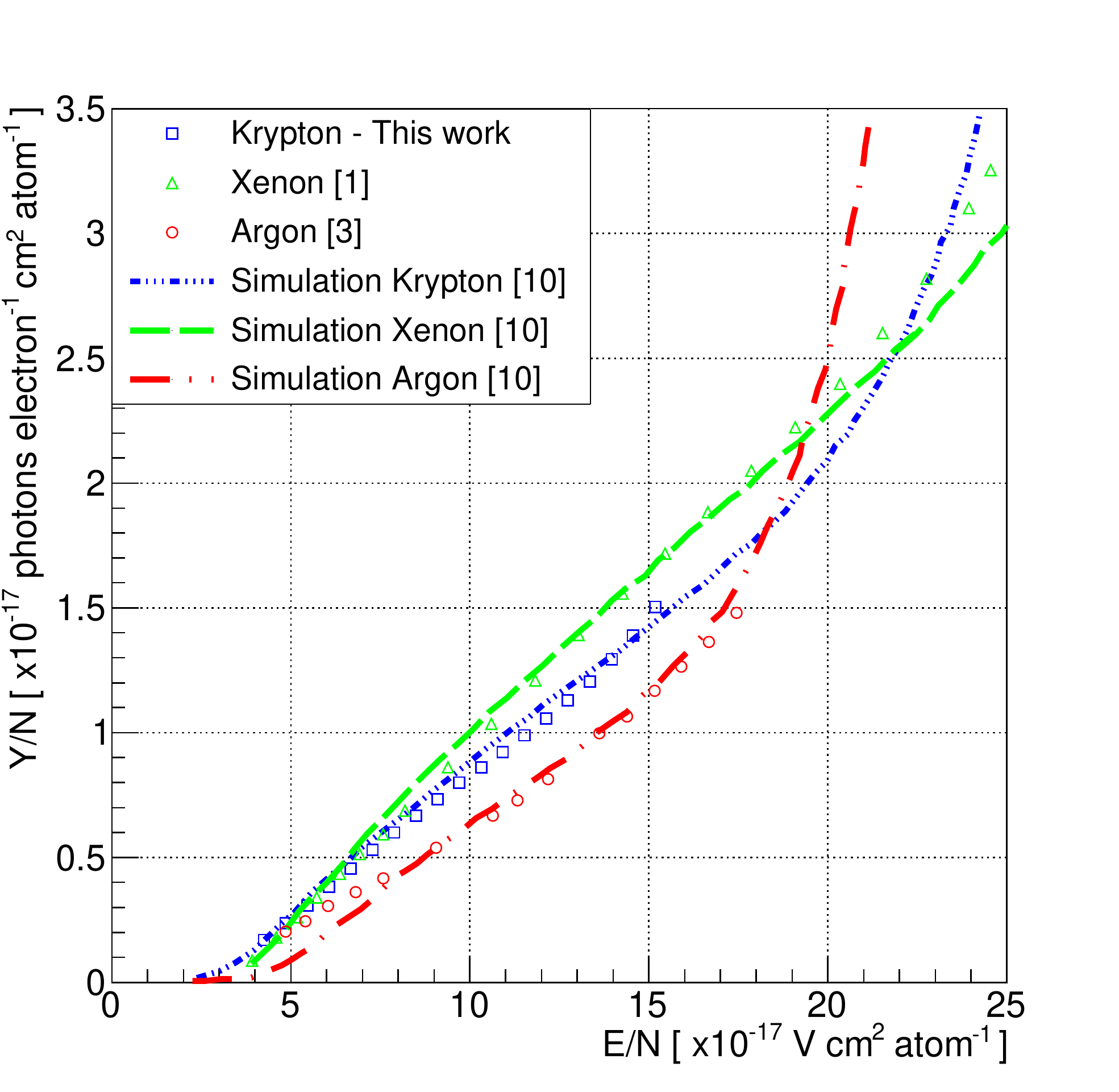}
    \caption{\label{Fig4} Reduced electroluminescence yield in Xe, Kr and Ar as a function of reduced electric field. Data points correspond to experimental data from \cite{1}, this work and \cite{3}, while the curves correspond to simulation data \cite{10}.}
\end{figure}

For comparison, Fig.\ref{Fig4} depicts the experimental results obtained with the present method for the electroluminescence yield in Xe \cite{1}, Kr (this work) and Ar \cite{3} along with the respective simulation results obtained by the most recent simulation work-package for EL production in noble gases \cite{10}. The scintillation amplification parameter in Kr and Ar is about 80$\%$ and 60$\%$, respectively, of that for Xe.

\section{Conclusions}
\label{sec:5 concl}

We have performed experimental studies on the reduced electroluminescence yield of pure Kr at room temperature and compared it with those obtained by Monte Carlo simulation. For the experimental measurements we used a gas proportional scintillation counter (GPSC), having a VUV-sensitive large area avalanche photodiode (LAAPD) for the scintillation readout. We used 5.9-keV x-rays to induce electroluminescence in the GPSC or to interact directly in the LAAPD. The concurrent detection of the electroluminescence pulses and the x-rays in the photosensor under the same conditions and in the same setup, allows to use the number of charge carriers produced by the x-rays interacting directly in the LAAPD as a reference for determining the number of charge carriers produced by the electroluminescence in the LAAPD, allowing a straightforward measurement of the number of photons impinging on it.
The reduced electroluminescence yield exhibits a linear dependence with reduced electric field, with an amplification parameter of 113 photons per kV, per electron, and a scintillation threshold of 2.7 Td (0.7 kV cm$^{-1}$ bar$^{-1}$ at 293 K), in good agreement with the simulation data present in the literature. Above 14 Td (3.5 kV cm$^{-1}$ bar$^{-1}$) the reduced electroluminescence yield departs from the linear trend, showing a faster increase due to the additional scintillation produced by extra secondary electrons resulting from the charge multiplication onset. The Kr amplification parameter is about 80$\%$ and 140$\%$ of that measured for Xe and Ar, respectively.

\

\

\

\acknowledgments

 This work is funded by FEDER, through the Programa Operacional Factores de Competitividade - COMPETE and by National funds through FCT - Fundação para a Ciência e Tecnologia in the frame of project PTDC/FIS/NUC/1534/2014 and UID/FIS/04559/2020 (LIBPhys).


\end{document}